# Bounds for the $l_1$-distance of $q$-ary lattices obtained via Constructions D, $\mathrm{D}'$ and $\overline{\mathrm{D}}$

**Eleonesio Strey** · **Sueli I. R. Costa**

**Abstract** Lattices have been used in several problems in coding theory and cryptography. In this paper we approach $q$-ary lattices obtained via Constructions D, $\mathrm{D}'$ and $\overline{\mathrm{D}}$. It is shown connections between Constructions D and $\mathrm{D}'$. Bounds for the minimum $l_1$-distance of lattices $\Lambda_D$, $\Lambda_{D'}$ and $\Lambda_{\overline{D}}$ and, under certain conditions, a generator matrix for $\Lambda_{D'}$ are presented. In addition, when the chain of codes used is closed under the zero-one addition, we derive explicit expressions for the minimum $l_1$-distances of the lattices $\Lambda_D$ and $\Lambda_{\overline{D}}$ attached to the distances of the codes used in these constructions.

## 1 Introduction

A lattice is a discrete additive subgroup of $\mathbb{R}^n$. A full rank lattice is the set of all integer linear combinations of a basis of $\mathbb{R}^n$. Lattices can be related to linear codes over finite rings and have been used for error correction in different contexts (see [21], [22] and their references). Recently, lattices have also been used in the proposal of cryptographic schemes (see [9], [10] and their references). We consider here $q \in \mathbb{N}$, the ring $\mathbb{Z}_q$ of integers modulo $q$ and linear codes $C \subseteq \mathbb{Z}_q^n$ (additive subgroups of $\mathbb{Z}_q^n$). There are several constructions involving linear codes and lattices, for

E. Strey
Department of Pure and Applied Mathematics,
Federal University of Espirito Santo, Brazil
E-mail: eleonesio.strey@ufes.br

S. I. R. Costa
Institute of Mathematics,
University of Campinas, Brazil
E-mail: sueli@ime.unicamp.br

This work was partially supported by FAPESP 2013/25997-7 and CNPq 312926/2013-8. Preliminary versions of some results presented in Section 3 were presented in [17].

example the Construction A [3] and the Constructions D, D′ and $\overline{\text{D}}$ [18]. These constructions appear in recent works such as [1], [4], [5], [6], [7], [12], [13], [14] and [20]. Given a lattice, a problem of interest is to determine its minimum distance in a given metric. In this article, we consider the $l_1$-distance (also known as Manhattan distance) and discuss the minimum $l_1$-distance of the lattices $\Lambda_D$, $\Lambda_{D'}$ and $\Lambda_{\overline{D}}$ obtained from Constructions D, D′ and $\overline{\text{D}}$, respectively. The paper is organized as follows. Concepts and preliminary results are introduced in Section 2. In Section 3, we present the Constructions D, D′ and $\overline{\text{D}}$ and also connections between them. The results on the minimum $l_1$-distances of lattices $\Lambda_D, \Lambda_{D'}$ and $\Lambda_{\overline{D}}$ are derived in Section 4. Concluding remarks are included in Section 5.

## 2 Preliminaries

In this Section we summarize the main concepts and results to be used in the next sections. We may quote [3] and [21] as general references.

A *q-ary linear code* $C$ of length $n$ over the ring $\mathbb{Z}_q$ is a $\mathbb{Z}_q$-submodule of $\mathbb{Z}_q^n$, that is, an additive subgroup of $\mathbb{Z}_q^n$. We denote by $C = \langle \boldsymbol{b}_1, ..., \boldsymbol{b}_k \rangle$ the code generated by the vectors $\boldsymbol{b}_1, ..., \boldsymbol{b}_k \in \mathbb{Z}_q^n$. A set $\{\boldsymbol{b}_1, ..., \boldsymbol{b}_k\}$ is a basis for $C$ when the vectors $\boldsymbol{b}_1, ..., \boldsymbol{b}_k$ are linearly independent and generate $C$. In this case, any element $\boldsymbol{v} \in C$ can be written in a unique way as a linear combination of the vectors $\boldsymbol{b}_1, ..., \boldsymbol{b}_k$. In opposition to codes over fields there are $q$-ary linear codes (if $q$ is not a prime) that have no basis, for example, the code $C = \langle(2,4)\rangle = \{(0,0),(2,4),(4,2)\} \subseteq \mathbb{Z}_6^2$ does not have a basis because every nonempty subset of $C$ is linearly dependent $(3(a,b) = (0,0))$. Every $q$-ary linear code $C$ is characterized by a minimal set of generators. A generator matrix for a $q$-ary linear code $C$ is any matrix whose rows form a minimal set of generators for the code $C$. For each pair of vectors $\boldsymbol{x} = (x_1, ..., x_n)$ and $\boldsymbol{y} = (y_1, ..., y_n)$ in $\mathbb{Z}_q^n$, the *inner product* between $\boldsymbol{x}$ and $\boldsymbol{y}$ is defined, as usual, as $\boldsymbol{x} \cdot \boldsymbol{y} = x_1 y_1 + \cdots + x_n y_n$. O set $C^\perp = \{\boldsymbol{y} \in \mathbb{Z}_q^n;\ \boldsymbol{x} \cdot \boldsymbol{y} = 0, \forall \boldsymbol{x} \in C\}$ is a $q$-ary linear code, called the *dual code* of $C$. If $C_1$ and $C_2$ are $q$-ary linear codes such that $C_1 \supseteq C_2$, then $C_1^\perp \subseteq C_2^\perp$.

A *lattice* $\Lambda$ is a discrete additive subgroup of $\mathbb{R}^n$. Equivalently, $\Lambda \subseteq \mathbb{R}^n$ is a lattice if and only if there are linearly independent vectors $\boldsymbol{v}_1, ..., \boldsymbol{v}_m \in \mathbb{R}^n$ such that $\Lambda$ is the set of all integer linear combinations of $\boldsymbol{v}_i$, $i = 1, ..., m$, i.e.,

$$\Lambda = \{\alpha_1 \boldsymbol{v}_1 + \cdots + \alpha_m \boldsymbol{v}_m;\ \alpha_1, ..., \alpha_m \in \mathbb{Z}\}.$$

In the above description, the set $\{\boldsymbol{v}_1, ..., \boldsymbol{v}_m\}$ is called a *basis* of $\Lambda$ and the number $m$ is called the *rank* of $\Lambda$. When $m = n$ we say that $\Lambda$ has *full rank.* The matrix $M$ whose rows are the vectors $\boldsymbol{v}_1, ..., \boldsymbol{v}_m$ is called a *generator matrix* of $\Lambda$. $M_1$ and $M_2$ are generator matrices of the same lattice if and only if there is a unimodular matrix $U$ (with integer entries and $\det(U) = \pm 1$) such that $M_2 = UM_1$. The *determinant* of $\Lambda$ is defined as $\det \Lambda = \det(MM^t)$ and this is an invariant under change of basis. The *volume* of $\Lambda$ is defined as $\text{vol}(\Lambda) = \sqrt{\det \Lambda}$, and corresponds to the Euclidean volume of parallelotope $P = \{\sum_{i=1}^n \alpha_i \boldsymbol{v}_i;\ 0 \leq \alpha_i < 1\}$, a fundamental region of $\Lambda$. We denote by $\text{span}(M) = \{\boldsymbol{u}M;\ \boldsymbol{u} \in \mathbb{R}^n\}$ the vector space generated by the rows of the matrix $M$. The *dual lattice* of $\Lambda = \Lambda(M)$, denoted by $\Lambda^* = \Lambda^*(M)$, is defined as

$$\Lambda^* = \{\boldsymbol{w} \in \text{span}(M);\ \langle \boldsymbol{v}, \boldsymbol{w} \rangle \in \mathbb{Z}, \forall \boldsymbol{v} \in \Lambda\},$$

where $\langle , \rangle$ is the canonical inner product in $\mathbb{R}^n$. It can be shown that $M$ is a generator matrix for $\Lambda$ iff $(MM^t)^{-1}M$ (pseudoinverse of $M^t$) is a generator matrix for $\Lambda^*$. The dual of $k\Lambda$ is $(1/k)\Lambda^*$, for all $0 \neq k \in \mathbb{R}$. For any lattice $\Lambda$, $(\Lambda^*)^* = \Lambda$.

A *sphere packing* in $\mathbb{R}^n$ is a collection of spheres/balls of equal size which do not overlap (except for touching). A *lattice packing* in $\mathbb{R}^n$ is a sphere packing in $\mathbb{R}^n$ such that the set formed by the centers of the spheres is a lattice. The *packing radius* $\rho$ of a lattice $\Lambda$ relative to a distance $d$ is the largest real number $r$ such that $\Lambda + B_d[\mathbf{0}, r]$ is a lattice packing, where $B_d[\mathbf{0}, r]$ is the closed ball with center $\mathbf{0}$ and radius $r$ relative to distance $d$. If translations by $\boldsymbol{v} \in \Lambda$ are isometries under $d$, it can be shown that $\rho = \lambda/2$, where $\lambda$ is the value of the minimum distance of $\Lambda$ relative to distance $d$, i.e.,

$$\lambda = \min\{d(\boldsymbol{x}, \boldsymbol{y});\ \boldsymbol{x}, \boldsymbol{y} \in \Lambda \text{ and } \boldsymbol{x} \neq \boldsymbol{y}\} = \min_{\mathbf{0} \neq \boldsymbol{x} \in \Lambda} d(\boldsymbol{x}, \mathbf{0}).$$

The *packing density* of a lattice $\Lambda$ of rank $n$ relative to a distance $d$ is defined as

$$\Delta_d(\Lambda) = \frac{\mathrm{vol}(B_d[\mathbf{0}, \rho])}{\mathrm{vol}(\Lambda)} = \frac{\mathrm{vol}(B_d[\mathbf{0}, 1])\rho^n}{\sqrt{\det \Lambda}},$$

where $\mathrm{vol}(B_d[\mathbf{0}, \rho])$ is the Euclidean volume of the ball $B_d[\mathbf{0}, \rho]$. The *center density* of a lattice $\Lambda$ of rank $n$ relative to distance $d$ is the number $\delta_d(\Lambda) = \rho^n/\mathrm{vol}(\Lambda)$. Given a distance $d$ in $\mathbb{R}^n$, a problem of interest is the search for a lattice with greatest possible density and such lattices are known only for a very few dimensions. For example, for the Euclidean distance, the densest lattices are known in the dimensions from 1 to 8 [3] and 24 [2]. For the $l_1$-distance, densest lattices are only known in dimensions $1, 2$ and 3 [5].

The $l_1$-*distance* (also known as *Manhattan distance*) between two elements $\boldsymbol{x}$ and $\boldsymbol{y}$ of $\mathbb{R}^n$ is defined as

$$d^1(\boldsymbol{x}, \boldsymbol{y}) = \|\boldsymbol{x} - \boldsymbol{y}\|_1 = \sum_{i=1}^{n} |x_i - y_i|.$$

For each lattice $\Lambda \subseteq \mathbb{R}^n$, the minimum $l_1$-distance of $\Lambda$ is defined as

$$d^1_{min}(\Lambda) = \min\{d^1(\boldsymbol{x}, \boldsymbol{y})|\ \boldsymbol{x}, \boldsymbol{y} \in \Lambda \text{ and } \boldsymbol{x} \neq \boldsymbol{y}\}.$$

It can be shown that if $d$ is the $l_1$-distance, then $\mathrm{vol}(B_d[\mathbf{0}, 1]) = 2^n/n!$ [11]. Given two elements $\boldsymbol{x}, \boldsymbol{y} \in \mathbb{Z}_q^n$, the *Lee distance* between $\boldsymbol{x}$ and $\boldsymbol{y}$ is defined as

$$d_{Lee}(\boldsymbol{x}, \boldsymbol{y}) = \sum_{i=1}^{n} \min\{\sigma(x_i - y_i), q - \sigma(x_i - y_i)\},$$

where $\sigma : \mathbb{Z}_q \to \mathbb{Z}$ is the standard inclusion map. The *minimum Lee distance* of a linear code $C \subseteq \mathbb{Z}_q^n$, $C \neq \{\mathbf{0}\}$, is defined as

$$d_{Lee}(C) = \min\{d_{Lee}(\boldsymbol{x}, \boldsymbol{y});\ \boldsymbol{x}, \boldsymbol{y} \in C \text{ and } \boldsymbol{x} \neq \boldsymbol{y}\}.$$

The Lee distance was introduced in [8] and [19] in the approach of signal transmission over certain channels with noise. Recent applications in the communications area can be found, for example, in [6] and its references.

## 3 Connections between Constructions D, D$'$ and $\overline{\mathrm{D}}$

Let $\sigma : \mathbb{Z}_q \to \mathbb{Z}$ be the standard inclusion map associated with the natural ring homomorphism $\overline{\sigma} : \mathbb{Z} \to \mathbb{Z}_q$, $\overline{\sigma}(x) = \overline{x}$, and the respective extensions $\sigma : \mathbb{Z}_q^n \to \mathbb{Z}^n$ and $\overline{\sigma} : \mathbb{Z}^n \to \mathbb{Z}_q^n$.

**Definition 1** (Construction A) Let $C \subseteq \mathbb{Z}_q^n$ be a linear code. The Construction A lattice associated with $C$, $\Lambda_A(C)$, is defined as

$$\Lambda_A(C) = \overline{\sigma}^{-1}(C) = q\mathbb{Z}^n + \sigma(C).$$

$\Lambda_A(C)$ is always a full rank lattice [21, p.31]. In [15] it is shown that given a linear code $\{\mathbf{0}\} \neq C \subseteq \mathbb{Z}_q^n$, the minimum $l_1$-distance of the lattice $\Lambda_A(C)$ is

$$d^1_{\min}\big(\Lambda_A(C)\big) = \min\big\{q, d_{Lee}(C)\big\}.$$

**Example 1.** Consider the linear code

$$C = \langle (1, 36, 3), (0, 37, 7) \rangle \subseteq \mathbb{Z}_{38}^3$$

and the lattice $\Lambda_A(C) = 38\mathbb{Z}^3 + \sigma(C)$. It is easy to see that

$$\begin{bmatrix} 1 & 36 & 3 \\ 0 & -1 & 7 \\ 0 & 0 & 38 \end{bmatrix}$$

is a generator matrix of $\Lambda_A(C)$. Thus, the matrix

$$M = \begin{bmatrix} 1 & 38 & -7 \\ -2 & -75 & 14 \\ 3 & 107 & -20 \end{bmatrix} \begin{bmatrix} 1 & 36 & 3 \\ 0 & -1 & 7 \\ 0 & 0 & 38 \end{bmatrix} = \begin{bmatrix} 1 & -2 & 3 \\ -2 & 3 & 1 \\ 3 & 1 & -2 \end{bmatrix}$$

is also a generator matrix of $\Lambda_A(C)$, because

$$\begin{bmatrix} 1 & 38 & -7 \\ -2 & -75 & 14 \\ 3 & 107 & -20 \end{bmatrix}$$

is an unimodular matrix. In [11] it was shown that the density of $\Lambda(M)$ (and hence of $\Lambda_A(C)$) relative to $l_1$-distance is 18/19 and that this is the highest density relative to $l_1$-distance in $\mathbb{R}^3$.

**Definition 2** (Construction D) Let $\mathbb{Z}_q^n \supseteq C_1 \supseteq C_2 \supseteq \cdots \supseteq C_a$ be a family of nested linear codes. Choose integers $k_1 \geq k_2 \geq \cdots \geq k_a \geq 0$ and vectors $\boldsymbol{b}_1, ..., \boldsymbol{b}_{k_1}$ in $\mathbb{Z}_q^n$ such that $\boldsymbol{b}_1, ..., \boldsymbol{b}_{k_\ell}$ span $C_\ell$, for $\ell = 1, ..., a$. The set $\Lambda_D$ consists of all vectors of the form

$$\boldsymbol{z} + \sum_{\ell=1}^{a} \sum_{j=1}^{k_\ell} \beta_j^{(\ell)} \frac{1}{q^{\ell-1}} \sigma(\boldsymbol{b}_j),$$

where $\boldsymbol{z} \in q\mathbb{Z}^n$ and $\beta_j^{(\ell)} \in \{0, 1, ..., q-1\}$.

It can be shown [18] that $\Lambda_D$ consists of all vectors of the form

$$q\boldsymbol{z} + \sum_{i=1}^{a} \sum_{j=k_{i+1}+1}^{k_i} \alpha_j^{(i)} \frac{1}{q^{i-1}} \sigma(\boldsymbol{b}_j),$$

where $\boldsymbol{z} \in \mathbb{Z}^n$ and $\alpha_j^{(i)} \in \{0, 1, ..., q^i - 1\}$.

**Definition 3** (Construction D$^{'}$) Let $\mathbb{Z}_q^n \supseteq C_1 \supseteq C_2 \supseteq \cdots \supseteq C_a$ be a family of nested linear codes. Choose integers $r_1, r_2, ..., r_a$ satisfying $0 \leq r_1 \leq r_2 \leq \cdots \leq r_a$ and vectors $\boldsymbol{h}_1, ..., \boldsymbol{h}_{r_a}$ in $\mathbb{Z}_q^n$ such that

$$C_\ell^\perp = \langle \boldsymbol{h}_1, ..., \boldsymbol{h}_{r_\ell} \rangle \quad \text{for} \quad \ell = 1, 2, ..., a$$

where $C_\ell^\perp$ is the dual code of $C_\ell$. We define $\Lambda_{D'}$ as the set consisting of all vectors $\boldsymbol{x} \in \mathbb{Z}^n$ satisfying the congruences

$$\boldsymbol{x} \cdot \sigma(\boldsymbol{h}_j) \equiv \boldsymbol{0} \mod q^{i+1}$$

for every pair $(i, j)$ satisfying $0 \leq i < a$ and $r_{a-i-1} < j \leq r_{a-i}$, where $r_0 = 0$.

In [18] it is shown that the sets $\Lambda_D$ and $\Lambda_{D'}$ respectively obtained via Constructions D and D$^{'}$ from $\mathbb{Z}_q^n \supseteq C_1 \supseteq C_2 \supseteq \cdots \supseteq C_a$ are always lattices of full rank in $\mathbb{R}^n$. When $a = 1$, we have $\Lambda_D = \Lambda_{D'} = \Lambda_A(C_1)$.

**Definition 4** (Construction $\overline{\text{D}}$) Let $\mathbb{Z}_q^n \supseteq C_1 \supseteq C_2 \supseteq \cdots \supseteq C_a$ be a family of nested linear codes. We define the set $\Gamma_{\overline{D}}$ as follows

$$\Gamma_{\overline{D}} = q^a \mathbb{Z}^n + q^{a-1} \sigma(C_1) + \cdots + q^1 \sigma(C_{a-1}) + \sigma(C_a).$$

When $a = 1$, we have $\Gamma_{\overline{D}} = \Lambda_A(C_1)$ and hence $\Gamma_{\overline{D}}$ is a lattice. For $a \geq 2$, we have $\Gamma_{\overline{D}}$ is a discrete set, but it is not always a lattice. The smallest lattice containing $\Gamma_{\overline{D}}$ is denoted by $\Lambda_{\overline{D}}$. In other words, $\Lambda_{\overline{D}}$ satisfies the following properties: (i) $\Lambda_{\overline{D}}$ is a lattice containing $\Gamma_{\overline{D}}$ and (ii) se $\Lambda \supseteq \Gamma_{\overline{D}}$ is a lattice, then $\Lambda \supseteq \Lambda_{\overline{D}}$.

For each pair of vectors $\boldsymbol{x} = (x_1, ..., x_n)$ and $\boldsymbol{y} = (y_1, ..., y_n)$ in $\mathbb{Z}_q^n$, the zero-one addition is defined as

$$\boldsymbol{x} * \boldsymbol{y} = (x_1 * y_1, ..., x_n * y_n) \in \mathbb{Z}_q^n,$$

where

$$x_i * y_i = \begin{cases} 0, & \text{if } \sigma(x_i) + \sigma(y_i) < q \\ 1, & \text{if } \sigma(x_i) + \sigma(y_i) \geq q. \end{cases}$$

We say that a family of nested linear codes $\mathbb{Z}_q^n \supseteq C_1 \supseteq C_2 \supseteq \cdots \supseteq C_a$ is closed under the zero-one addition if and only if the zero-one addition of any two elements of $C_i$ is contained in $C_{i-1}$, for $i = 2, ..., a$. For instance, the chain $\mathbb{Z}_3^2 \supseteq C_1 \supseteq C_2$, where $C_1 = C_2 = \langle (1, 2) \rangle = \{(0, 0), (1, 2), (2, 1)\}$, is not closed under the zero-one addition, since $(1, 2) \in C_2$ and $(1, 2) * (1, 2) = (0, 1) \notin C_1$. In [18] it is shown that $\Gamma_{\overline{D}}$ is a lattice if and only if the chain $\mathbb{Z}_q^n \supseteq C_1 \supseteq C_2 \supseteq \cdots \supseteq C_a$ is closed under the zero-one addition. In this case, $\Lambda_{\overline{D}} = \Gamma_{\overline{D}} = q^{a-1} \Lambda_D$.

In the next two lemmas we extend to q-ary codes results previously known for binary codes [20].

**Lemma 1** *Let $H$ be the $k_1 \times n$ matrix, whose rows are the vectors*

$$\sigma(\boldsymbol{b}_1), ..., \sigma(\boldsymbol{b}_{k_a}), q\sigma(\boldsymbol{b}_{k_a+1}), ..., q\sigma(\boldsymbol{b}_{k_{a-1}}), ..., q^{a-1}\sigma(\boldsymbol{b}_{k_2+1}), ..., q^{a-1}\sigma(\boldsymbol{b}_{k_1}).$$

*Then, $\boldsymbol{x} \in \Lambda_D^*$ if and only if $q\boldsymbol{x} \in \mathbb{Z}^n$ and $H\boldsymbol{x}^t \equiv \boldsymbol{0} \mod q^{a-1}$.*

*Proof.* By definition, $\Lambda_D^* = \{\boldsymbol{x} \in \mathbb{R}^n; \langle \boldsymbol{x}, \boldsymbol{y} \rangle \in \mathbb{Z}, \forall \boldsymbol{y} \in \Lambda_D\}$. Thus $\boldsymbol{x} \in \Lambda_D^*$ if and only if $\langle \boldsymbol{x}, q\boldsymbol{z} \rangle \in \mathbb{Z}$ and $\left\langle \boldsymbol{x}, (1/q^{i-1})\sigma(\boldsymbol{b}_j) \right\rangle \in \mathbb{Z}$ for $1 \leq i \leq a$, $k_{i+1} < j \leq k_i$ and for all $\boldsymbol{z} \in \mathbb{Z}^n$. That is, $\boldsymbol{x} \in \Lambda_D^*$ if and only if

$$q^{a-i} \langle \boldsymbol{x}, \sigma(\boldsymbol{b}_j) \rangle \equiv q^a \langle \boldsymbol{x}, \boldsymbol{e}_t \rangle \equiv 0 \mod q^{a-1}$$

for $1 \leq i \leq a$, $k_{i+1} < j \leq k_i$ and $1 \leq t \leq n$, where $\{\boldsymbol{e}_1, ..., \boldsymbol{e}_n\}$ is the standard base of $\mathbb{R}^n$. Therefore $\boldsymbol{x} \in \Lambda_D^*$ if and only if $q\boldsymbol{x} \in \mathbb{Z}^n$ and $H\boldsymbol{x}^t \equiv \boldsymbol{0} \mod q^{a-1}$. □

**Lemma 2** *Let $H$ be the $r_a \times n$ matrix, whose rows are the vectors*

$$\sigma(\boldsymbol{h}_1), ..., \sigma(\boldsymbol{h}_{r_1}), q\sigma(\boldsymbol{h}_{r_1+1}), ..., q\sigma(\boldsymbol{h}_{r_2}), ..., q^{a-1}\sigma(\boldsymbol{h}_{r_{a-1}+1}), ..., q^{a-1}\sigma(\boldsymbol{h}_{r_a}).$$

*Then, $\boldsymbol{x} \in \Lambda_{D'}$ if and only if $\boldsymbol{x} \in \mathbb{Z}^n$ and $H\boldsymbol{x}^t \equiv \boldsymbol{0} \mod q^a$.*

*Proof.* It is sufficient to observe that $\boldsymbol{x} \in \Lambda_{D'}$ if and only if $\boldsymbol{x} \in \mathbb{Z}^n$ and $q^{a-i}\boldsymbol{x} \cdot \sigma(\boldsymbol{h}_j) \equiv \boldsymbol{0} \mod q^a$ for $0 < i \leq a$ and $r_{a-i} < j \leq r_{a-i+1}$. Therefore $\boldsymbol{x} \in \Lambda_{D'}$ if and only if $\boldsymbol{x} \in \mathbb{Z}^n$ and $H\boldsymbol{x}^t \equiv \boldsymbol{0} \mod q^a$. □

Given a chain of linear codes $\mathbb{Z}_q^n \supseteq C_1 \supseteq C_2 \supseteq \cdots \supseteq C_a$ and parameters $0 \leq r_1 \leq r_2 \leq \cdots \leq r_a$ and $\boldsymbol{h}_1, ..., \boldsymbol{h}_{r_a} \in \mathbb{Z}_q^n$ such that $C_i^\perp = \langle \boldsymbol{h}_1, ..., \boldsymbol{h}_{r_i} \rangle$ for $i = 1, 2, ..., a$, let $\Lambda_{D'}$ be the lattice obtained via Construction D$'$ from this chain (using the parameters $r_1, r_2, ..., r_a, \boldsymbol{h}_1, ..., \boldsymbol{h}_{r_a}$), ie,

$$\Lambda_{D'} = \left\{ \boldsymbol{x} \in \mathbb{Z}^n \middle| \ \boldsymbol{x} \cdot \sigma(\boldsymbol{h}_j) \equiv \boldsymbol{0} \mod q^{i+1}, 0 \leq i < a \text{ and } r_{a-i-1} < j \leq r_{a-i} \right\},$$

where $r_0 = 0$. Let $\Lambda_{D^\perp}$ be the lattice obtained via Construction D from the chain $C_1^\perp \subseteq C_2^\perp \subseteq \cdots \subseteq C_a^\perp \subseteq \mathbb{Z}_q^n$ (using the parameters $r_1, r_2, ..., r_a, \boldsymbol{h}_1, ..., \boldsymbol{h}_{r_a}$), ie, the set $\Lambda_{D^\perp}$ consists of all vectors of the form

$$q\boldsymbol{z} + \sum_{i=1}^{a} \sum_{j=r_{a-i}+1}^{r_{a-i+1}} \alpha_j^{(i)} \frac{1}{q^{i-1}} \sigma(\boldsymbol{h}_j)$$

where $\boldsymbol{z} \in \mathbb{Z}^n$ and $\alpha_j^{(i)} \in \{0, 1, ..., q^i - 1\}$. The following results consider the above notations.

**Theorem 1** *Let $\mathbb{Z}_q^n \supseteq C_1 \supseteq C_2 \supseteq \cdots \supseteq C_a$ be a family of nested linear codes. Then $q\Lambda_{D^\perp}^* = \Lambda_{D'}$. Moreover, $M$ is a generator matrix of $\Lambda_{D^\perp}$ if and only if $q(M^t)^{-1}$ is a generator matrix of $\Lambda_{D'}$.*

*Proof.* Let $H$ be the matrix, whose rows are the vectors

$$\sigma(\boldsymbol{h}_1), ..., \sigma(\boldsymbol{h}_{r_1}), q\sigma(\boldsymbol{h}_{r_1+1}), ..., q\sigma(\boldsymbol{h}_{r_2}), ..., q^{a-1}\sigma(\boldsymbol{h}_{r_{a-1}+1}), ..., q^{a-1}\sigma(\boldsymbol{h}_{r_a}).$$

Then,

$$\begin{aligned}
\boldsymbol{y} \in q\Lambda^*_{D^\perp} &\overset{\text{Lemma 1}}{\Longleftrightarrow} \boldsymbol{y} = q\boldsymbol{x}, q\boldsymbol{x} \in \mathbb{Z}^n \text{ and } H\boldsymbol{x}^t \equiv \boldsymbol{0} \mod q^{a-1} \\
&\Longleftrightarrow \boldsymbol{y} \in \mathbb{Z}^n \text{ and } H\boldsymbol{y}^t \equiv \boldsymbol{0} \mod q^a \\
&\overset{\text{Lemma 2}}{\Longleftrightarrow} \boldsymbol{y} \in \Lambda_{D'}.
\end{aligned}$$

Thus $q\Lambda^*_{D^\perp} = \Lambda_{D'}$. We note that $M$ is a generator matrix of $\Lambda_{D^\perp}$ if and only if $(M^t)^{-1}$ is a generator matrix of $\Lambda^*_{D^\perp}$. Therefore $M$ is a generator matrix of $\Lambda_{D^\perp}$ if and only if $q(M^t)^{-1}$ is a generator matrix of $\Lambda_{D'}$, since $\Lambda_{D'} = q\Lambda^*_{D^\perp}$. □

**Corollary 1** *The chain $C_1^\perp \subseteq C_2^\perp \subseteq \cdots \subseteq C_a^\perp \subseteq \mathbb{Z}_q^n$ is closed under the zero-one addition if and only if $\Lambda_{D'} = q^a \Gamma^*_{\overline{D}^\perp} = q^a \Lambda^*_{\overline{D}^\perp}$, onde $\Gamma_{\overline{D}^\perp}$ and $\Lambda_{\overline{D}^\perp}$ are obtained via Construction $\overline{D}$ from the chain $C_1^\perp \subseteq C_2^\perp \subseteq \cdots \subseteq C_a^\perp \subseteq \mathbb{Z}_q^n$.*

*Proof.* The chain $C_1^\perp \subseteq C_2^\perp \subseteq \cdots \subseteq C_a^\perp \subseteq \mathbb{Z}_q^n$ is closed under zero-one addition if and only if $\Lambda_{\overline{D}^\perp} = \Gamma_{\overline{D}^\perp} = q^{a-1}\Lambda_{D^\perp}$ [18]. On the other hand, $\Lambda_{\overline{D}^\perp} = \Gamma_{\overline{D}^\perp} = q^{a-1}\Lambda_{D^\perp}$ if and only if $\Gamma^*_{\overline{D}^\perp} = \Lambda^*_{\overline{D}^\perp} = (q^{a-1}\Lambda_{D^\perp})^* = (1/q^{a-1})\Lambda^*_{D^\perp} = (1/q^a)\Lambda_{D'}$. So the chain $C_1^\perp \subseteq C_2^\perp \subseteq \cdots \subseteq C_a^\perp \subseteq \mathbb{Z}_q^n$ is closed under zero-one addition if and only if $\Lambda_{D'} = q^a \Gamma^*_{\overline{D}^\perp} = q^a \Lambda^*_{\overline{D}^\perp}$. □

The next result obtained from Theorem 1 when $a = 1$ is known [10].

**Corollary 2** *Let $C \subseteq \mathbb{Z}_q^n$ be a linear code. If $\Lambda_A(C)$ and $\Lambda_A(C^\perp)$ are the lattices obtained via Construction A from $C$ and $C^\perp$ respectively, where $C^\perp$ is the dual code of $C$, then*

$$\Lambda^*_A(C) = \frac{1}{q}\Lambda_A(C^\perp).$$

The next theorem provides, under certain conditions, a generator matrix and the determinant of the lattice $\Lambda_{D'}$ depending on the parameters used in its construction.

**Theorem 2** *Let $\boldsymbol{h}_1, ..., \boldsymbol{h}_{r_a} \in \mathbb{Z}_q^n$ be nonzero vectors such that*

1. *$C_\ell^\perp = \langle \boldsymbol{h}_1, ..., \boldsymbol{h}_{r_\ell} \rangle$ for $\ell = 0, 1, ..., a$.*
2. *Some row permutation of matrix $[\sigma(\boldsymbol{h}_1) \ \cdots \ \sigma(\boldsymbol{h}_{r_a})]^t$ forms an upper triangular (respectively, lower triangular) matrix in the row echelon form.*
3. *For each $j \in \{1, ..., r_a\}$, the first (respectively the last) nonzero component of vector $\sigma(\boldsymbol{h}_j)$, denoted by $\alpha_j$, divides $q$ as well as all other components of this vector.*

*Then there is a upper triangular (respectively, lower triangular) generator matrix $M$ of $\Lambda_D^\perp$ whose rows are the $r_a$ vectors $(1/q^{i-1})\sigma(\boldsymbol{h}_j)$, where $1 \le i \le a$ and $r_{a-i} < j \le r_{a-i+1}$, plus $n - r_a$ vectors of the shape $(0, ..., 0, q, 0, ..., 0)$. Consequently,*

$$\det \Lambda_D^\perp = \det(MM^t) = \left(\prod_{j=1}^{r_a} \alpha_j\right)^2 \left(q^2\right)^{n - \sum_{\ell=1}^{a} r_\ell}.$$

*Furthermore, $q(M^t)^{-1}$ is a generator matrix of $\Lambda_{D'}$ and hence*

$$\det \Lambda_{D'} = \left(\prod_{j=1}^{r_a} \alpha_j\right)^{-2} \left(q^2\right)^{\sum_{\ell=1}^{a} r_\ell}.$$

*Proof.* The first part of this theorem is a consequence of [18,Theorem 6]. To conclude this proof it is sufficient observe that, by Theorem 1, $q(M^t)^{-1}$ is a generator matrix of $\Lambda_{D'}$. Therefore

$$\det \Lambda_{D'} = \det [q(M^t)^{-1}(q(M^t)^{-1})^t] = (q^2)^n \det (MM^t)^{-1} = (q^2)^n (\det \Lambda_{D^\perp})^{-1}.$$

□

**Example 2.** Let $\mathbb{Z}_6^2 \supseteq C_1 \supseteq C_2$ be the chain of linear codes, where $C_1 = \langle(1,2)\rangle$ and $C_2 = \langle(2,4)\rangle$. We have $C_1^\perp = \{(x,y) \in \mathbb{Z}_6^2; x + 2y = 0\} = \langle(4,1)\rangle$ and $C_2^\perp = \{(x,y) \in \mathbb{Z}_6^2; 2x + 4y = 0\} = \langle(4,1),(3,0)\rangle$. Let $\Lambda_{D^\perp}$ be the lattice obtained via Construction D from this chain using the parameters $r_1 = 1, r_2 = 2, \boldsymbol{h}_1 = (4,1)$ and $\boldsymbol{h}_2 = (3,0)$. These parameters satisfy the conditions of Theorem 2, thus

$$\det \Lambda_{D^\perp} = \left(3 \cdot 1\right)^2 \left(6^2\right)^{2-(1+2)} = 1/4$$

and

$$\det \Lambda_{D'} = \left(3 \cdot 1\right)^{-2} \left(6^2\right)^{1+2} = (6^3/3)^2 = 72^2.$$

In addition, generator matrices for $\Lambda_{D^\perp}$ and $\Lambda_{D'}$ are given respectively by

$$M = \begin{pmatrix} \sigma(\boldsymbol{h}_2) \\ (1/6)\sigma(\boldsymbol{h}_1) \end{pmatrix} = \begin{pmatrix} 3 & 0 \\ 2/3 & 1/6 \end{pmatrix} \text{ and } 6(M^t)^{-1} = \begin{pmatrix} 2 & -8 \\ 0 & 36 \end{pmatrix}.$$

Applying Theorem 2, we obtain $\Lambda_{D'}^* = (1/6)\Lambda_{D^\perp}$ and consequently a generator matrix for $\Lambda_{D'}^*$ is given by

$$\begin{pmatrix} 1/2 & 0 \\ 1/9 & 1/36 \end{pmatrix}.$$

## 4 Bounds for the minimum $l_1$-distances in $\Lambda_D, \Lambda_{D'}$ and $\Lambda_{\overline{D}}$

In this section, we present results regarding the minimum $l_1$-distance of the lattices $\Lambda_D, \Lambda_{D'}$ and $\Lambda_{\overline{D}}$. Very preliminary and abridged versions of some of these results were presented in [16].

**Lemma 3** *Let $\{\mathbf{0}\} \neq C \subseteq \mathbb{Z}_q^n$ be a linear code. There are $\boldsymbol{x}, \boldsymbol{y} \in C$ such that*

$$\|\sigma(\boldsymbol{x}) - \sigma(\boldsymbol{y})\|_1 = d_{Lee}(C).$$

*Proof.* Let $\boldsymbol{z} \in C$ such that $d_{Lee}(\boldsymbol{z}, \mathbf{0}) = d_{Lee}(C)$. Thus $2\boldsymbol{z} \in C$ and $\sigma(2\boldsymbol{z}) - \sigma(\boldsymbol{z}) = (\alpha_1, ..., \alpha_n)$, where

$$\alpha_i = \begin{cases} \sigma(z_i) & \text{se } \sigma(z_i) < q/2 \\ \sigma(z_i) - q, & \text{se } \sigma(z_i) \geq q/2. \end{cases}$$

Therefore

$$\|\sigma(2\boldsymbol{z}) - \sigma(\boldsymbol{z})\|_1 = \sum_{i=1}^n |\alpha_i| = \sum_{i=1}^n \min\{\sigma(z_i), q - \sigma(z_i)\} = d_{Lee}(\boldsymbol{z}, \mathbf{0}) = d_{Lee}(C).$$

This shows that there are $\boldsymbol{x}, \boldsymbol{y} \in C$ such that $\|\sigma(\boldsymbol{x}) - \sigma(\boldsymbol{y})\|_1 = d_{Lee}(C)$. □

**Theorem 3** *Let $\mathbb{Z}_q^n \supseteq C_1 \supseteq C_2 \supseteq \cdots \supseteq C_a \neq \{\mathbf{0}\}$ be a family of nested linear codes and let $\Gamma_{\overline{D}}$ be the set obtained via Construction $\overline{D}$ from this family. If the minimum Lee distance in $C_\ell$ is $d_{Lee}^{\ell}$, for $\ell = 1, 2, ...a$, then*

$$d_{\min}^1(\Gamma_{\overline{D}}) = \min\{q^a, q^{a-1}d_{Lee}^1, ..., d_{Lee}^a\}.$$

*Proof.* For each $\ell \in \{1, 2, ..., a\}$, Lemma 3 assures that there are $\boldsymbol{x}_\ell, \boldsymbol{y}_\ell \in C_\ell$ such that $\|\sigma(\boldsymbol{x}_\ell) - \sigma(\boldsymbol{y}_\ell)\|_1 = d_{Lee}^{\ell}$. Since $q^{a-\ell}\sigma(C_\ell) \subseteq \Gamma_{\overline{D}}$, we have

$$q^{a-\ell}\sigma(\boldsymbol{x}_\ell), q^{a-\ell}\sigma(\boldsymbol{y}_\ell) \in \Gamma_{\overline{D}}$$

and

$$\|q^{a-\ell}\sigma(\boldsymbol{x}_\ell) - q^{a-\ell}\sigma(\boldsymbol{y}_\ell)\|_1 = q^{a-\ell}\|\sigma(\boldsymbol{x}_\ell) - \sigma(\boldsymbol{y}_\ell)\|_1 = q^{a-\ell}d_{Lee}^{\ell}.$$

Besides, $d_{\min}^1(\Gamma_{\overline{D}}) \leq q^a$ since $q^a\mathbb{Z}^n \subseteq \Gamma_{\overline{D}}$. Thus

$$d_{\min}^1(\Gamma_{\overline{D}}) \leq \min\{q^a, q^{a-1}d_{Lee}^1, ..., d_{Lee}^a\}.$$

To complete the proof, we will show that $d_{\min}^1(\Gamma_{\overline{D}}) \geq \min\{q^a, q^{a-1}d_{Lee}^1, ..., d_{Lee}^a\}$. Indeed, given $\boldsymbol{x}, \boldsymbol{y} \in \Gamma_{\overline{D}}$, we write $\boldsymbol{x} = q^\ell \boldsymbol{v}$ and $\boldsymbol{y} = q^k \boldsymbol{w}$ with $\boldsymbol{v}, \boldsymbol{w} \in \mathbb{Z}^n$, $\boldsymbol{v} \not\equiv \mathbf{0}$ mod $q$ and $\boldsymbol{w} \not\equiv \mathbf{0}$ mod $q$. We can assume without loss of generality that $\ell \geq k$. We remark that: (i) If $k \geq a$ then

$$d^1(\boldsymbol{x}, \boldsymbol{y}) = q^k d^1(q^{\ell-k}\boldsymbol{v}, \boldsymbol{w}) \geq q^a, \text{ because } \mathbf{0} \neq q^{\ell-k}\boldsymbol{v} - \boldsymbol{w} \in \mathbb{Z}^n.$$

(ii) If $0 \leq k \leq a-1$ and $0 \leq \ell \leq a-1$, then there are $\boldsymbol{c}_1 \in C_1, ..., \boldsymbol{c}_{a-k} \in C_{a-k}$ and $\boldsymbol{z} \in \mathbb{Z}^n$ such that $\boldsymbol{y} = q^a\boldsymbol{z} + q^{a-1}\sigma(\boldsymbol{c}_1) + \cdots + q^k\sigma(\boldsymbol{c}_{a-k})$ and consequently

$$\boldsymbol{w} = q^{a-k}\boldsymbol{z} + q^{a-1-k}\sigma(\boldsymbol{c}_1) + \cdots + q^0\sigma(\boldsymbol{c}_{a-k}).$$

Note that $\mathbf{0} \neq q^{\ell-k}\overline{\boldsymbol{v}} - \overline{\boldsymbol{w}} \in C_{a-k}$ (because $\overline{\boldsymbol{v}} \in C_{a-\ell}$ and $\overline{\boldsymbol{w}} \in C_{a-k}$) and

$$\begin{aligned} d^1(\boldsymbol{x}, \boldsymbol{y}) &= q^k d^1(q^{\ell-k}\boldsymbol{v}, \boldsymbol{w}) = q^k \sum_{i=1}^{n} |q^{\ell-k}v_i - w_i| \\ &\geq q^k \sum_{i=1}^{n} \min\{\sigma(q^{\ell-k}\overline{v}_i - \overline{w}_i), q - \sigma(q^{\ell-k}\overline{v}_i - \overline{w}_i)\} \geq q^k d_{Lee}^{a-k}, \end{aligned}$$

since $|q^{\ell-k}v_i - w_i| \geq \min\{\sigma(q^{\ell-k}\overline{v}_i - \overline{w}_i), q - \sigma(q^{\ell-k}\overline{v}_i - \overline{w}_i)\}$ for $i = 1, ..., n$.
(iii) If $0 \leq k \leq a-1$ and $\ell \geq a$, then $\mathbf{0} \neq \overline{\boldsymbol{w}} \in C_{a-k}$ and $\boldsymbol{x} = q^a\boldsymbol{z}$, where $\boldsymbol{z} = q^{\ell-a}\boldsymbol{v} \in \mathbb{Z}^n$. Thus,

$$\begin{aligned} d^1(\boldsymbol{x}, \boldsymbol{y}) &= q^k d^1(q^{a-k}\boldsymbol{z}, \boldsymbol{w}) = q^k \sum_{i=1}^{n} |q^{a-k}z_i - w_i| \\ &\geq q^k \sum_{i=1}^{n} \min\{\sigma(-\overline{w}_i), q - \sigma(-\overline{w}_i)\} \\ &= q^k d_{Lee}(-\overline{\boldsymbol{w}}, \mathbf{0}) \geq q^k d_{Lee}^{a-k}. \end{aligned}$$

Therefore $d^1(\boldsymbol{x}, \boldsymbol{y}) \geq \min\{q^a, q^{a-1}d_{Lee}^1, ..., d_{Lee}^a\}, \forall \boldsymbol{x}, \boldsymbol{y} \in \Gamma_{\overline{D}}$. □

**Corollary 3** *[15] Let* $\{\mathbf{0}\} \neq C \subseteq \mathbb{Z}_q^n$ *be a linear code and let* $\Lambda_A(C)$ *be the lattice obtained from* $C$ *via Construction A. Then,*

$$d_{\min}^1\big(\Lambda_A(C)\big) = \min\big\{q, d_{Lee}(C)\big\}.$$

**Corollary 4** *Let* $\mathbb{Z}_q^n \supseteq C_1 \supseteq C_2 \supseteq \cdots \supseteq C_a \neq \{\mathbf{0}\}$ *be a family of nested linear codes,* $\Gamma_{\overline{D}}$ *be the set obtained via Construction* $\overline{D}$ *from this family and let* $\Lambda_{\overline{D}}$ *be the smallest lattice containing* $\Gamma_{\overline{D}}$*. If the minimum Lee distance in* $C_\ell$ *is* $d_{Lee}^\ell$*, for* $\ell = 1, 2, ...a$*, then*

$$d_{\min}^1(\Lambda_{\overline{D}}) \leq \min\{q^a, q^{a-1}d_{Lee}^1, ..., d_{Lee}^a\}.$$

*When the family of nested linear codes is closed under the zero-one addition (ie,* $\Lambda_{\overline{D}} = \Gamma_{\overline{D}}$*), we have*

$$d_{\min}^1(\Lambda_{\overline{D}}) = \min\{q^a, q^{a-1}d_{Lee}^1, ..., d_{Lee}^a\}.$$

We believe that the second part of Corollary 4 can be refined by withdrawing the hypothesis that the chain $\mathbb{Z}_q^n \supseteq C_1 \supseteq C_2 \supseteq \cdots \supseteq C_a$ is closed under the zero-one addition. Therefore, we propose the following conjecture.

*A conjecture regarding Corollary 4:* Let $\mathbb{Z}_q^n \supseteq C_1 \supseteq C_2 \supseteq \cdots \supseteq C_a \neq \{\mathbf{0}\}$ be a family of nested linear codes, let $\Gamma_{\overline{D}}$ be the set obtained via Construction $\overline{\text{D}}$ from this family and let $\Lambda_{\overline{D}}$ be the smallest lattice containing $\Gamma_{\overline{D}}$. If the minimum Lee distance in $C_\ell$ is $d_{Lee}^\ell$, for $\ell = 1, 2, ...a$, then

$$d_{\min}^1(\Lambda_{\overline{D}}) = \min\{q^a, q^{a-1}d_{Lee}^1, ..., d_{Lee}^a\}.$$

**Corollary 5** *Let* $\mathbb{Z}_q^n \supseteq C_1 \supseteq C_2 \supseteq \cdots \supseteq C_a \neq \{\mathbf{0}\}$ *be a family of nested linear codes closed under the zero-one addition and let* $\Lambda_D$ *be the lattice obtained via Construction D from this family. If the minimum Lee distance in* $C_\ell$ *is* $d_{Lee}^\ell$*, for* $\ell = 1, 2, ...a$*, then*

$$d_{\min}^1(\Lambda_D) = \min_{1\leq \ell \leq a}\left\{q, \frac{1}{q^{\ell-1}}d_{Lee}^\ell\right\}.$$

*Proof.* Since the chain of codes used is closed under the zero-one addition, we have $q^{a-1}\Lambda_D = \Lambda_{\overline{D}}$ and hence $d_{\min}^1(\Lambda_D) = (1/q^{a-1})\min\{q^a, q^{a-1}d_{Lee}^1, ..., d_{Lee}^a\}$. □

In the next example we show that the hypothesis of the chain of linear codes used in Construction D is closed under the zero-one addition can not be omitted in the Corollary 5.

**Example 3.** Let $\mathbb{Z}_3^2 \supseteq C_1 \supseteq C_2$ be a family of nested linear codes, where $C_1 = C_2 = \langle(1, 2)\rangle$. Choosing the parameters $k_1 = 2, k_2 = 1$ and $\boldsymbol{b}_1 = (1, 2), \boldsymbol{b}_2 = (2, 1) \in \mathbb{Z}_3^2$, we have $0 \leq k_2 \leq k_1$, $C_1 = \langle \boldsymbol{b}_1, \boldsymbol{b}_2\rangle$, $C_2 = \langle \boldsymbol{b}_1\rangle$ and consequently $\Lambda_D$ consists of all vectors of the form

$$\boldsymbol{z} + \alpha_2^{(1)}(2, 1) + \alpha_1^{(2)}\frac{1}{3}(1, 2),$$

where $\boldsymbol{z} \in 3\mathbb{Z}^2, 0 \leq \alpha_2^{(1)} < 3$ and $0 \leq \alpha_1^{(2)} < 9$. Note that $3\Lambda_D$ is a 9-ary lattice. The elements of $(3\Lambda_D) \cap [0, 9)^2$ are shown in Figure 1. In this example, we observe

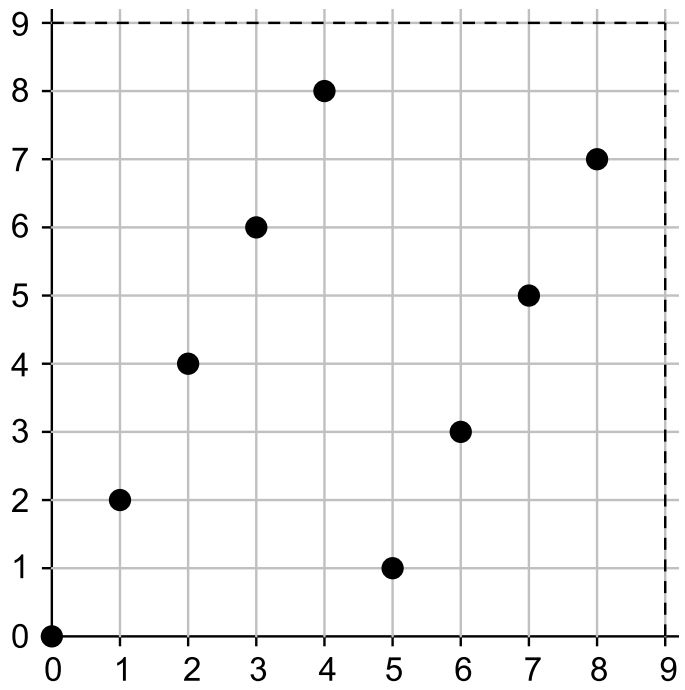

**Fig. 1** The elements of $3\Lambda_D$ inside the box $[0,9)^2$.

that $d^1_{Lee} = d^2_{Lee} = 2$, $d^1_{\min}(\Lambda_D) = 1$,

$$\min_{1\leq \ell \leq 2}\left\{3, \frac{1}{3^{\ell-1}} d^{\ell}_{Lee}\right\} = \frac{2}{3}$$

and that the chain $C_1 \supseteq C_2$ is not closed under zero-one addition, since $(1,2) \in C_2$ e $(1,2) * (1,2) = (0,1) \notin C_1$.

We remark that Example 3 does not contradict the conjecture regarding Corollary 4. Indeed, we have $\Gamma_{\overline{D}} = 3^2\mathbb{Z}^2 + 3^1\sigma(C_1) + 3^0\sigma(C_2)$ where $\sigma(C_1) = \sigma(C_2) = \{(0,0),(1,2),(2,1)\}$, i.e., $\Gamma_{\overline{D}} = 3^2\mathbb{Z}^2 + \Gamma_{\overline{D}} \cap [0,9)^2$ and

$$\Gamma_{\overline{D}} \cap [0,9)^2 = \{(0,0),(3,6),(6,3),(1,2),(4,8),(7,5),(2,1),(5,7),(8,4)\}.$$

The elements of $\Gamma_{\overline{D}} \cap [0,9)^2$ are shown in Figure 2(a). Note that set $\Gamma_{\overline{D}}$ is not a lattice (since the chain of codes is not closed under the zero-one addition). In the figure 2(b), we present the lattice $\Lambda_{\overline{D}}$ (that is, the smallest lattice that contains $\Gamma_{\overline{D}}$). We note that $d^1_{\min}(\Lambda_{\overline{D}}) = 2$ (see Figure 2(b)) and $\min\{3^2, 3d^1_{Lee}, d^2_{Lee}\} = \min\{9,6,2\} = 2$. It shows that the Example 3 is in accordance with the Conjecture **??**.

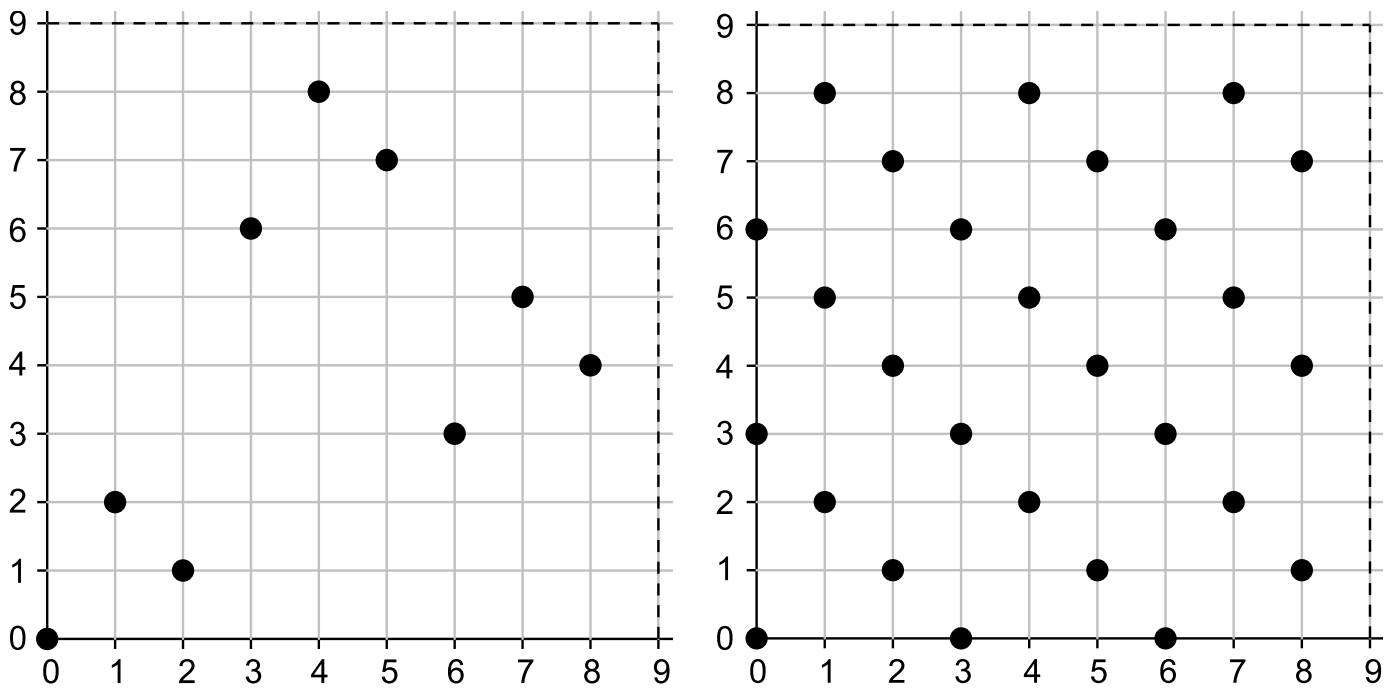

(a) $\Gamma_{\overline{D}} \cap [0,9)^2$ (b) $\Lambda_{\overline{D}} \cap [0,9)^2$

**Fig. 2** The elements of $\Gamma_{\overline{D}}$ and $\Lambda_{\overline{D}}$ inside the box $[0,9)^2$

**Corollary 6** *Let $\mathbb{Z}_q^n \supseteq C_1 \supseteq C_2 \supseteq \cdots \supseteq C_a$ be a family of nested linear codes and let $\Lambda_{D'}$ be the lattice obtained via Construction $D'$ from this family. If $C_1^\perp \neq \{\mathbf{0}\}$ and the family $\mathbb{Z}_q^n \supseteq C_a^\perp \supseteq \cdots \supseteq C_2^\perp \supseteq C_1^\perp$ is closed under the zero-one addition, then the minimum $l_1$-distance of $\Lambda_{D'}^*$ (dual lattice of $\Lambda_{D'}$) is given by*

$$d_{\min}^1(\Lambda_{D'}^*) = \min_{1\leq \ell \leq a} \left\{1, \frac{d_{Lee}^{\ell,\perp}}{q^{a-\ell+1}}\right\}.$$

*where $d_{Lee}^{\ell,\perp}$ is the minimum Lee distance in $C_\ell^\perp$, for $\ell = 1, 2, ...a$.*

*Proof.* It follows immediately from Theorem 1 along with Corollary 5. □

**Corollary 7** *Let $\mathbb{Z}_q^n \supseteq C_1 \supseteq C_2 \supseteq \cdots \supseteq C_a \neq \{\mathbf{0}\}$ be a family of nested linear codes closed under the zero-one addition. If the minimum Lee distance $d_{Lee}^\ell$ of $C_\ell$ satisfies $d_{Lee}^\ell \geq q^\ell$, $\ell = 1, 2, ...a$, then $d_{\min}^1(\Lambda_{\overline{D}}) = q^a$ and $d_{\min}^1(\Lambda_D) = q$.*

**Corollary 8** *Let $\mathbb{Z}_q^n \supseteq C_1 \supseteq C_2 \supseteq \cdots \supseteq C_a \neq \{\mathbf{0}\}$ be a family of nested linear codes closed under the zero-one addition. Let us assume that the vectors $\boldsymbol{b}_1, ..., \boldsymbol{b}_{k_1} \in \mathbb{Z}_q^n$ used in Construction D are nonzero and satisfy:*

1. *Some row permutation of matrix $[\sigma(\boldsymbol{b}_1)\cdots\sigma(\boldsymbol{b}_{k_1})]^t$ forms an upper triangular (respectively, lower triangular) matrix in the row echelon form.*
2. *For each $j \in \{1, ..., k_1\}$, the first (respectively, the last) nonzero component of vector $\sigma(\boldsymbol{b}_j)$, denoted by $\alpha_j$, divides $q$ as well as all other components of this vector.*

*If $\lambda = \min_{1\leq\ell\leq a}\{q, (1/q^{\ell-1})d_{Lee}^\ell\}$, where $d_{Lee}^\ell$ is the minimum Lee distance in $C_\ell$, $\ell = 1, 2, ...a$, the density of $\Lambda_D$ relative to $l_1$-distance is given by*

$$\Delta_1(\Lambda_{\overline{D}}) = \Delta_1(\Lambda_D) = \frac{\lambda^n q^{\sum_{\ell=1}^a k_\ell - n}}{n! \prod_{i=1}^{k_1} \alpha_i}$$

*and center density*

$$\delta_1(\Lambda_{\overline{D}}) = \delta_1(\Lambda_D) = \frac{\lambda^n q^{\sum_{\ell=1}^a k_\ell - n}}{2^n \prod_{i=1}^{k_1} \alpha_i}.$$

*Furthermore, $d_{Lee}^\ell \geq q^\ell$ if and only if*

$$\Delta_1(\Lambda_D) = \frac{q^{\sum_{\ell=1}^a k_\ell}}{n! \prod_{i=1}^{k_1} \alpha_i} \quad \text{and} \quad \delta_1(\Lambda_D) = \frac{q^{\sum_{\ell=1}^a k_\ell}}{2^n \prod_{i=1}^{k_1} \alpha_i}.$$

*Proof.* It follows from Theorem 3 and [18, Teorema 6] since the Euclidean volume of the ball $B_1[\mathbf{0}, 1]$ relative to $l_1$-distance is given by $\text{vol}(B_1[\mathbf{0}, 1]) = 2^n/n!$. □

The next theorem extends to $q$-ary codes results previously known for binary codes [20].

**Theorem 4** *Let $\mathbb{Z}_q^n \supseteq C_1 \supseteq C_2 \supseteq \cdots \supseteq C_a \neq \{\mathbf{0}\}$ be a family of nested linear codes, $0 \leq r_1 \leq r_2 \leq \cdots \leq r_a$ and $\boldsymbol{h}_1, ..., \boldsymbol{h}_{r_a} \in \mathbb{Z}_q^n$ such that $C_\ell^\perp = \langle \boldsymbol{h}_1, ..., \boldsymbol{h}_{r_\ell}\rangle$ for $\ell = 1, 2, ..., a$. If $\Lambda_{D'}$ is the lattice obtained via Construction $D'$ using the parameters described above, then*

$$\min\{q^a, q^{a-1}d_{Lee}^1, ..., d_{Lee}^a\} \leq d_{\min}^1(\Lambda_{D'}) \leq q^a,$$

*where $d_{Lee}^\ell$ is the minimum Lee distance in $C_\ell$, for $\ell = 1, 2, ...a$.*

*Proof.* If $\boldsymbol{x} \in \mathbb{Z}^n$ is a vector with a component which is not a multiple of $q$ and $\boldsymbol{x} \cdot \sigma(\boldsymbol{h}_j) \equiv \boldsymbol{0} \mod q$, for $1 \leq j \leq r_k$, then $\|\boldsymbol{x}\|_1 \geq d_{Lee}^k$. Indeed, we can write $\boldsymbol{x} = \boldsymbol{c} + q\boldsymbol{z}$, where $\boldsymbol{z} \in \mathbb{Z}^n$, $\boldsymbol{c} = (c_1, ..., c_n)$ and $c_i \in \{0, 1, ..., q-1\}$, $i = 1, 2, ..., n$. Since $\boldsymbol{x}$ has a component that is not a multiple of $q$, we have $\boldsymbol{c} \neq \boldsymbol{0}$. Besides, $\boldsymbol{c} \cdot \sigma(\boldsymbol{h}_j) \equiv \boldsymbol{0} \mod q$, $1 \leq j \leq r_k$, thus $\overline{\boldsymbol{c}} \in C_k$ and consequently $\|\boldsymbol{c}\|_1 \geq d_{Lee}^k$. Therefore

$$\|\boldsymbol{x}\|_1 = \|\boldsymbol{c} + q\boldsymbol{z}\|_1 = \sum_{i=1}^{n} |c_i + qz_i| \geq \sum_{i=1}^{n} \min\{c_i, q - c_i\} \geq d_{Lee}^k.$$

Now, let $\boldsymbol{0} \neq \boldsymbol{x} \in \Lambda_{D'}$. Since $\boldsymbol{x} \neq \boldsymbol{0}$, there is $k \geq 0$ such that $q^{-k}\boldsymbol{x} \in \mathbb{Z}^n$ and $q^{-k-1}\boldsymbol{x} \notin \mathbb{Z}^n$. If $k < a$ then $q^{-k}\boldsymbol{x} \cdot \sigma(\boldsymbol{h}_j) \equiv \boldsymbol{0} \mod q$ for $1 \leq j \leq r_{a-k}$, since

$$\begin{aligned}
\boldsymbol{x} \in \Lambda_{D'} &\iff \boldsymbol{x} \cdot \sigma(\boldsymbol{h}_j) \equiv \boldsymbol{0} \mod q^{i+1}, \text{ for } 0 \leq i < a \text{ and } r_{a-i-1} < j \leq r_{a-i}. \\
&\implies \boldsymbol{x} \cdot \sigma(\boldsymbol{h}_j) \equiv \boldsymbol{0} \mod q^{k+1}, \text{ for } k \leq i < a \text{ and } r_{a-i-1} < j \leq r_{a-i}. \\
&\iff q^{-k}\boldsymbol{x} \cdot \sigma(\boldsymbol{h}_j) \equiv \boldsymbol{0} \mod q, \text{ for } 1 < j \leq r_{a-k}.
\end{aligned}$$

Thus, $\|\boldsymbol{x}\|_1 \geq q^k d_{Lee}^{a-k}$. If $k \geq a$ then $\boldsymbol{x} = q^k\boldsymbol{z} = q^a(q^{k-a}\boldsymbol{z})$, for some $\boldsymbol{0} \neq \boldsymbol{z} \in \mathbb{Z}^n$, and so $\|\boldsymbol{x}\|_1 \geq q^a$. Therefore $\min\{q^a, q^{a-1}d_{Lee}^1, ..., d_{Lee}^a\} \leq d_{\min}^1(\Lambda_{D'})$. To obtain the inequality $d_{\min}^1(\Lambda_{D'}) \leq q^a$, it is sufficient to observe that $q^a\mathbb{Z}^n \subseteq \Lambda_{D'}$. □

**Example 4.** Let $\mathbb{Z}_3^2 \supseteq C_1 \supseteq C_2$ be the chain of linear codes, where $C_1 = C_2 = \langle(1,1)\rangle$. We have $C_1^\perp = C_2^\perp = \{(x,y) \in \mathbb{Z}_3^2; x + y = 0\} = \{(0,0),(1,2),(2,1)\} = \langle(1,2)\rangle$. Thus, for $r_1 = 1$, $r_2 = 2$ and $\boldsymbol{h}_1 = (1,2), \boldsymbol{h}_2 = (1,2) \in \mathbb{Z}_3^2$, we have $0 \leq r_1 \leq r_2$, $C_1^\perp = \langle\boldsymbol{h}_1\rangle$, $C_2^\perp = \langle\boldsymbol{h}_1, \boldsymbol{h}_2\rangle$ and consequently,

$$\Lambda_{D'} = \left\{(x,y) \in \mathbb{Z}^2 \middle|\ x + 2y \equiv 0 \mod 9\right\}.$$

Therefore

$$\Lambda_{D'} = \bigcup_{\boldsymbol{z} \in 9\mathbb{Z}^2} \left(\boldsymbol{z} + \Lambda_{D'} \cap [0,9)^2\right),$$

where the elements of $\Lambda_{D'} \cap [0,9)^2$ are shown in Figure 3. In this example, $d_{Lee}^1 = d_{Lee}^2 = 2$ and hence $\min\{3^2, 3d_{Lee}^1, d_{Lee}^2\} = 2 < 3 = d_{\min}^1(\Lambda_D') < 3^2$.

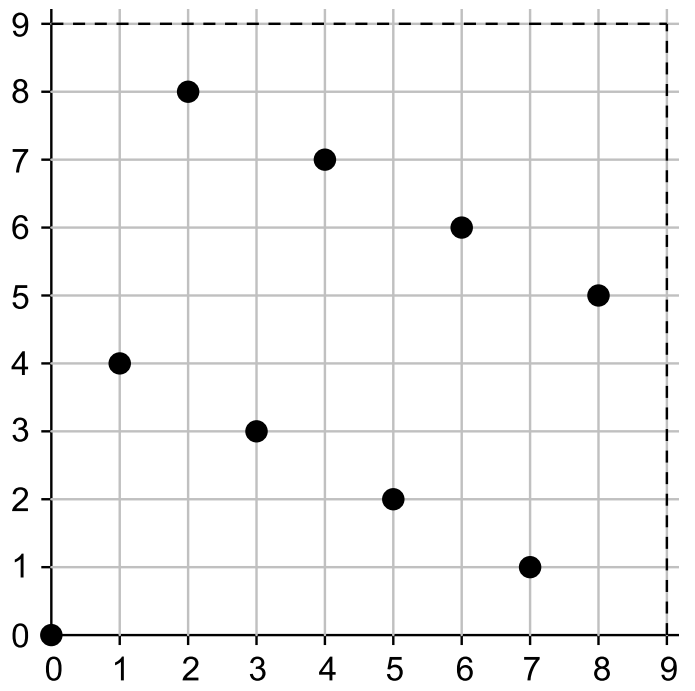

**Fig. 3** The elements of $\Lambda_{D'}$ inside the box $[0,9)^2$.

**Example 5.** An example where the inequality of Theorem 4 is strict can be given by considering the chain $\mathbb{Z}_6^2 \supseteq C_1 \supseteq C_2$, where $C_1 = \langle(1,2)\rangle$ and $C_2 = \langle(2,4)\rangle$ along with the parameters $r_1 = 1$, $r_2 = 2$ and $\boldsymbol{h}_1 = (4,1), \boldsymbol{h}_2 = (3,0) \in \mathbb{Z}_6^2$. In this example, it can be shown that $\min\{6^2, 3d_{Lee}^1, d_{Lee}^2\} = 4 < 5 = d_{\min}^1(\Lambda_D') < 6^2$.

We remark that when the vectors $\boldsymbol{h}_1, ..., \boldsymbol{h}_{r_a}$ used in Construction D$'$ satisfy the hypotheses of Theorem 2, we can obtain the density of $\Lambda_{D'}$ relative to $l_1$-distance from Theorems 2 and 4.

## 5 Conclusion

The minimum $l_1$-distance of lattices $\Lambda_D$, $\Lambda_{D'}$ and $\Lambda_{\overline{D}}$ obtained via Constructions D, D$'$ e $\overline{\text{D}}$ is investigated here. We provide upper bounds for the minimum $l_1$-distance at these lattices. When the chain of codes used is closed under the zero-one addition, we derive explicit expressions for the minimum $l_1$-distances of the lattices $\Lambda_D$ and $\Lambda_{\overline{D}}$ depending on the Lee-distances of the codes used in these constructions. In addition, we also show connections between Constructions D and D$'$ and obtain under certain conditions the minimum $l_1$-distance of the dual lattice $\Lambda_{D'}^*$. Further work in the directions presented here is the discussion of the minimum distance of these lattices regarding the $l_p$-norm.